\documentclass[letterpaper, 10 pt, conference]{ieeeconf}  

\IEEEoverridecommandlockouts                              

\usepackage{xcolor}
\definecolor{revised}{RGB}{0,0,0}
\newcommand{\revise}[1]{\textcolor{revised}{#1}}

\usepackage{graphicx}
\usepackage{url}
\usepackage{booktabs} 
\usepackage{array}  
\usepackage{balance}

\title{\LARGE \bf
SocializeChat: A GPT-Based AAC Tool Grounded in Personal Memories to Support Social Communication
}

\author{Wei Xiang$^{1}$, Yunkai Xu$^{2}$, Yuyang Fang$^{1}$, Zhuyu Teng$^{1}$, Zhaoqu Jiang$^{1}$, Beijia Hu$^{1}$, and Jinguo Yang$^{1}$%
\thanks{$^{1}$Wei Xiang, Yuyang Fang, Zhuyu Teng, Zhaoqu Jiang, Beijia Hu, and Jinguo Yang are with Zhejiang University, Hangzhou, China. {\tt\small \{wxiang, fangyuyang, tzhuyu, zhaoqujiang, 3210104942\}@zju.edu.cn}, {\tt\small gouchengouceq@163.com}}%
\thanks{$^{2}$Yunkai Xu is currently a Master's student at Penn State University, USA. This work was completed while he was an undergraduate student at Zhejiang University. {\tt\small yqx5322@psu.edu}}%
}

\begin{document}

\maketitle
\thispagestyle{empty}
\pagestyle{empty}

\begin{abstract}
Elderly people with speech impairments often face challenges in engaging in meaningful social communication, particularly when using Augmentative and Alternative Communication (AAC) tools that primarily address basic needs. Moreover, effective chats often rely on personal memories, which is hard to extract and reuse. We introduce SocializeChat, an AAC tool that generates sentence suggestions by drawing on users’ personal memory records. By incorporating topic preference and interpersonal closeness, the system reuses past experience and tailors suggestions to different social contexts and conversation partners. SocializeChat not only leverages past experiences to support interaction, but also treats conversations as opportunities to create new memories, fostering a dynamic cycle between memory and communication. A user study shows its potential to enhance the inclusivity and relevance of AAC-supported social interaction.

\end{abstract}
\section{INTRODUCTION}

Augmentative and Alternative Communication (AAC) tools have been developed to address communication challenges experienced by elderly adults with speech impairment~\cite{hustad2020augmentative}. Despite their strong desire for social communication, they still face substantial barriers when using traditional AAC tools~\cite{sun2023survey}. Conversations with others tend to focus on physiological needs rather than social or emotional topics~\cite{davidson2003identifying}. This functional focus, coupled with the user's passive role and limited ability to provide frequent verbal responses, reduces opportunities for engaging in rich, meaningful conversations~\cite{havstam2011taking}. 

Such limitations not only narrow the content of social interaction but also diminish the potential to share personal memories, which is an essential component of maintaining relational bonds~\cite{guan2022does}. Users of these systems generally communicate at a significantly slower pace~\cite{seale2020interaction}, which hinders the flow of conversation and discourages extended social interaction. Besides, these tools typically rely on pre-set buttons~\cite{hustad2020augmentative}, often lack the expressive flexibility required for the memory-based communication. As a result, many AAC users eventually withdraw from any social communication~\cite{bircanin2019challenges}, leading to increased social isolation and emotional frustration~\cite{heidari2021prevalence}.

\revise{Recent developments in large language models (LLMs) have introduced new possibilities for sentence generation in AAC tools~\cite{vargas2020design, shen2022kwickchat}.} These innovations provide AAC users with the opportunity to engage in conversations at a pace similar to that of their non-disabled partners. However, appropriate social communication should be aligned with users' topic preferences and closeness with the people they are ``talking'' to, fostering interpersonal relationships~\cite{svennevig2000getting, chan2020intimacy}. The content generated by AI-supported AAC remains largely consistent regardless of the conversation partners~\cite{hustad2020augmentative}, but the communication aimed at social purposes changes with interpersonal elements~\cite{dai2022designing}. For instance, when one talks to people with different closeness, the depth of their conversations varies~\cite{feng2022effect}. User studies also have reported that the language generated by AI could make close partners feel insulted and damage the relationship~\cite{valencia2023less}.  Consequently, AAC users need to constantly modify the generated sentence during a conversation, which may result in tiredness considering their motor impairment~\cite{dai2022designing}.



\revise{We present \textit{SocializeChat}, an AAC system that suggests sentences based on users’ memories (e.g. familiar events, routines, and relationships, etc,) and social context. A user study highlights its potential to support meaningful social interactions.}


\section{Related Works}
\subsection{Social Communication with AAC}

\revise{Social communication supports relationship-building~\cite{garzaniti2011building} and is essential to rehabilitation for individuals with speech impairments~\cite{last2022exploring}, yet current AAC tools often fall short. Users face difficulties expressing emotions and maintaining relationships, affecting their social well-being~\cite{bucki2019emotional, dai2022designing}. Most AAC systems rely on static vocabularies~\cite{hustad2020augmentative}, limiting depth and diversity. However, many seek broader communication beyond clinical or familial settings~\cite{curtis2022state}, calling for AAC designs that reflect users’ social identities and relationships~\cite{dai2022designing}.}


\revise{Recent AAC research has explored AI-driven approaches that generate suggestions based on user history or context~\cite{vargas2020design, shen2022kwickchat}. Although such systems improve communication speed, they often produce generic responses that overlook interpersonal nuance~\cite{valencia2023less}. \textit{Fang et al.} incorporated memory-informed responses aligned with social strategies, but their system lacked systematic retrieval of user records and limited user control over AI-assisted participation~\cite{fang2023socialize}. Efforts to enhance the expressivity of memories through visual elements and voice settings have shown promise~\cite{fiannaca2018voicesetting, albert2022aid}, but these features are usually applied uniformly without accounting for the dynamics of different relationships~\cite{lafrance2003contingent}. This points to a critical opportunity: designing AAC tools that not only diversify modes of expression but also adapt dynamically to social context and user preferences.}

\subsection{Influential Dimensions in Social Communication}

\revise{\textbf{Topic preference} reflects what people expect to discuss with different partners. Aligned topics help build relationships~\cite{svennevig2000getting}, and preferences vary by gender, familiarity, and context~\cite{lafrance2003contingent}. Prior works also show that communication intent~\cite{jannach2021survey}, long-term interests~\cite{zhang2022multiple}, and personality~\cite{pal2023affects} also shape topic selection.}

\revise{\textbf{Closeness} affects both topic choice and disclosure depth~\cite{feng2022effect, rocca1998relationship}, influencing speech quantity, emotional expression and other linguistic strategies~\cite{morman2002changing, pei2020quantifying}. Moreover, the intimate communication itself can strengthen social bonds~\cite{chan2020intimacy}.}




\subsection{Memory in Social Communication}
In the Human-Centered Interaction (HCI) field, memory has long been explored as a rich site for social interaction. A large body of work has focused on designing technologies that help older adults recall and share memories by incorporating mixed media—such as photographs~\cite{wang2024promoting}, audio recordings~\cite{niema2016interactive}, and multi-sensory experiences~\cite{chan2021pablo, wei2023bridging}. These memory sharing systems often serve to strengthen intergenerational connections, elicit storytelling, and shared conversations among others. While these designs often aim to evoke individual memories, they also underscore a crucial insight: memory is not purely personal but socially situated. As Halbwachs argues, individual recollection is shaped and given meaning within collective and conversational contexts~\cite{halbwachs2020collective}. This perspective suggests a promising, yet underexplored, opportunity in AAC design: leveraging users’ own memories to support and sustain meaningful social interactions.



\section{System Construction}
\begin{figure}[htbp]
    \centering
    \includegraphics[width=1\linewidth]{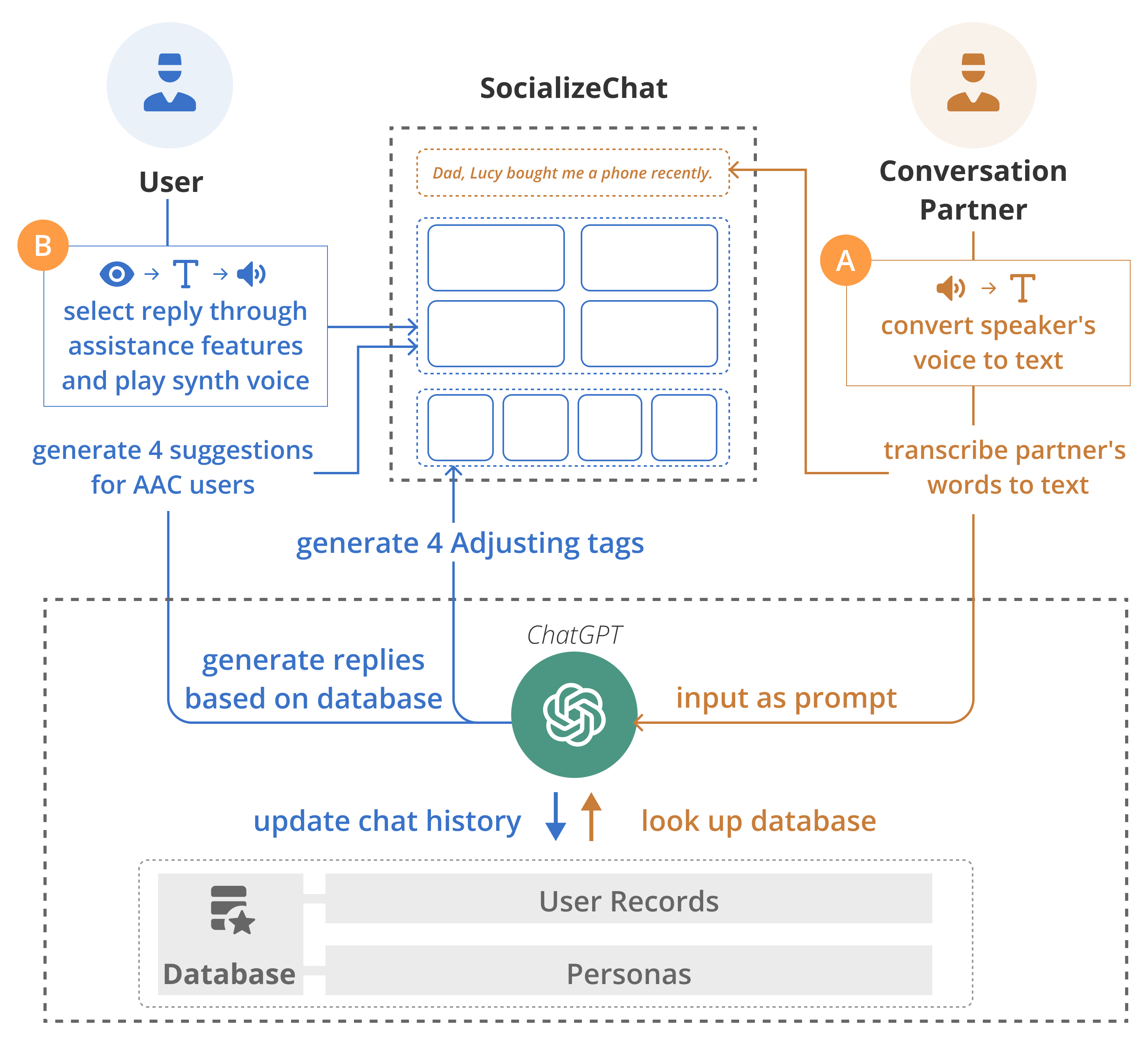}
    \caption{Usage Flow: When the conversation partner \revise{(the person communicating with the user)} is speaking to user (A). \revise{Voice input is transcribed and combined with retrieved memory records and partner persona as LLM prompt input; adjusted tags are generated alongside the response.} User\revise{(the AAC user operating SocializeChat) } can choose one of the provided sentence suggestions to engage in conversation or start a new conversation (B). }
    \label{fig:using-tool}
\end{figure}

SocializeChat is an AAC tool powered by ChatGPT-4, designed to support elderly users with speech impairments by leveraging the database (their memory records and personas of others) to facilitate communication, while also treating each conversation as a new memory for future use (Figure \ref{fig:using-tool}). 

\subsection{User Flow and Interface Design}
SocializeChat requires initial input to construct a user database, which consists of two main components that are user-editable: the personas of the conversation partners and user memory records (concrete text-based memories of users' events, routines and experience). The personas reflect two major dimensions of social communication: topic preferences and closeness. The topic preference is input by the user, indicating the topics that the user expects to discuss with the conversation partner, such as ``weather'' or ``grandson's studies''. Closeness is divided into three levels from which the user can choose one (comprehensive explanations for these three levels will be provided in the subsequent section \textit{Prompt Design}). User records are in text format and serve as source material when generating sentence suggestions.

\begin{figure}
    \centering
    \includegraphics[width=1\linewidth]{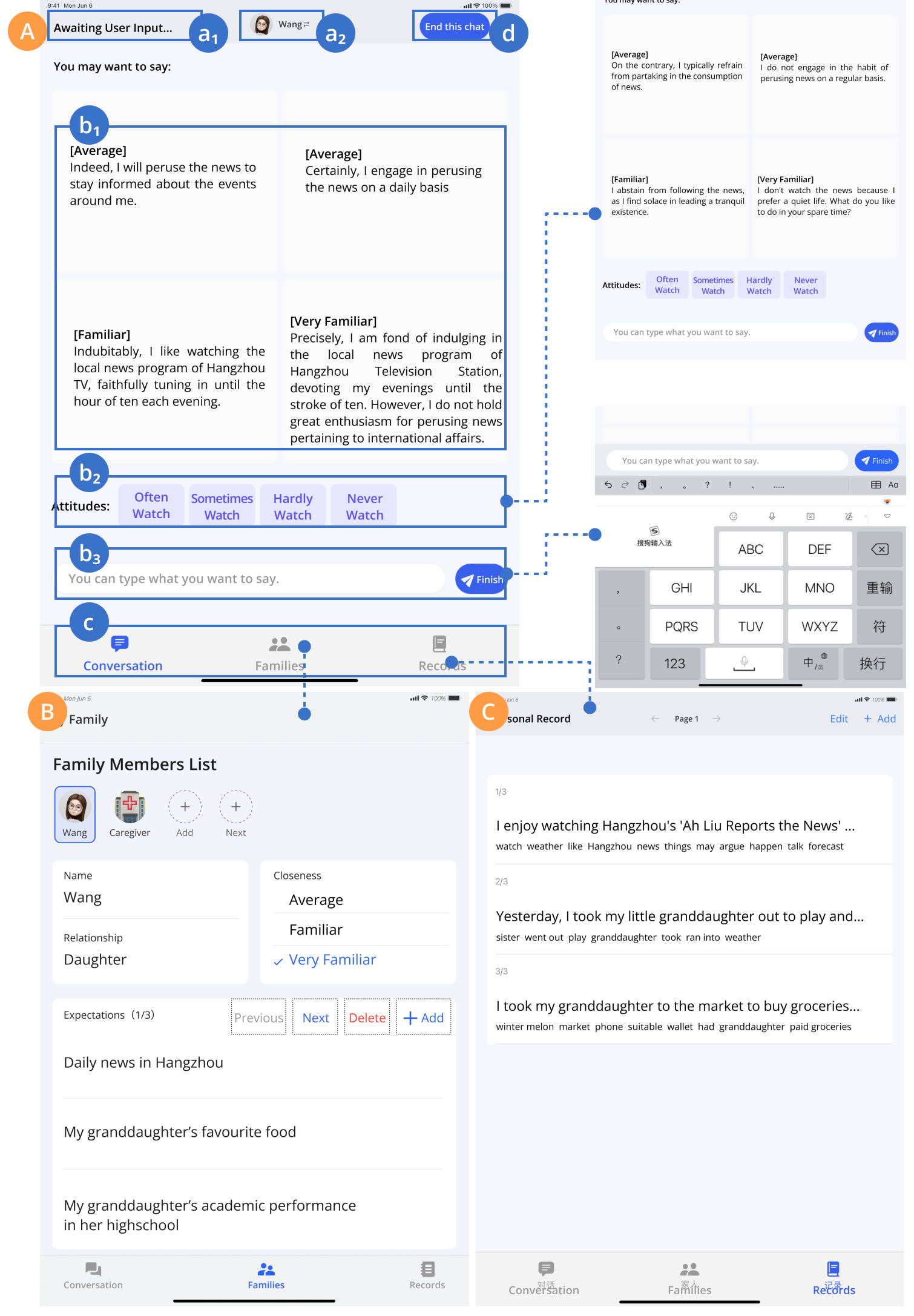}
    \caption{Interface Design: The main interface is divided into three sections: (A) system status and user identity ($a_1$, $a_2$); (B) the conversation interaction area, including sentence suggestions tagged with closeness levels ($b_1$), contextual tag adjustments ($b_2$), and a virtual keyboard for manual input ($b_3$); and (C) a navigation menu ($c$). Users can also initiate new topics based on partner preferences via a dedicated button ($d$).}
    \label{fig:user_inter}
\end{figure}

Before initiating conversations, the database can be filled out in the Families tab (Figure \ref{fig:user_inter} \textit{B}) and Records tab (Figure \ref{fig:user_inter} \textit{C}) respectively. Figure \ref{fig:user_inter} \textit{A} is the main interface of conversations.

\subsection{Sentence Generation Based on User Memory Records}
We start by providing \revise{two scenarios} to give an intuitive understanding. Then, we detail how response options are generated and influenced by various information.

The first scenario \revise{(S1)} described a conversation started by users. \revise{Before the conversation begins, the system selects up to four personal records from the user’s topic history with specific partners, offering a set of varied and relevant themes to initiate the dialogue.}


\revise{The second scenario \revise{(S2)} illustrates how relevant user records are retrieved during conversations to support context-aware responses. For example, when a partner says, ``I went to the park the day before yesterday… Would you like to go out and see it?''  , the system identifies a related memory such as ``I like fishing with friends in the park… watching the stars near XiShan Park.'' Based on this, GPT generates suggestions at varying levels of social closeness using our designed prompts:}


\begin{itemize}
  \item \textbf{Average:} ``Sure, I really like parks.''
  \item \textbf{Familiar:} ``Sure, I used to love stargazing in the park.''
  \item \textbf{Familiar:} ``Sure, I often used to fish with old friends in the park. I miss those times.''
  \item \textbf{Very Familiar:} ``Sure, I remember often visiting a temple in XiShan Park. I really like that place.''
\end{itemize}

\revise{The four responses convey similar meanings, with an initial tendency toward positive replies. Two of the responses are at the same level to offer users greater diversity.} The main differences among the options stem from Social Closeness, the description and specific differences of which can be found in the Section~\ref{sec:prompt-for-generation}.


\subsubsection{Preprocessing of Current Conversation with User Records}
\revise{Before using LLM to generate sentence suggestions, we analyze the current conversation and retrieve relevant user records. The process involves the following steps:}

\textbf{Step 1:}
Following speech recognition, the Apple NLP framework's API~\footnote{\url{https://developer.apple.com/documentation/naturallanguage/identifying_parts_of_speech}} is employed to extract keywords from the ongoing conversation.

\textbf{Step 2:}
For each keyword obtained, we apply Apple's Word Embedding API\footnote{\url{https://developer.apple.com/documentation/naturallanguage/nlembedding}} to map the keyword text to an embedding vector, which represents the meaning and contextual relationships of a word. The distance between different embedding vectors, such as the cosine distance we used, is positively correlated with the similarity between the corresponding word meanings. Through the embedding algorithm, we obtain 40 relevant words for each keyword (the quantity here controls the level of fuzziness in matching). The collection of all relevant words corresponding to all keywords is gathered to form a vocabulary set, serving as the basis for matching.

\textbf{Step 3:}
Calculation of priority for each record uses the vocabulary set obtained in step 2. For each record, the priority is calculated as follows (let the vocabulary set obtained in step 2 be \textit{A}, and the current vocabulary set is \textit{B}): Extract words from \textit{A}, and if a word exists in \textit{B}, the priority of that record is increased by one unit. This operation is performed for each word in \textit{A}, ultimately yielding a priority score for each record. \revise{The records are then ranked by their scores, with higher-ranked ones considered more contextually relevant.}


\textbf{Step 4:}
Three high-priority items from the record set are extracted and used as input to prompt, supplementing the sentence suggestions generated by GPT with \revise{personal memory}.

\subsubsection{Prompts for Generation}
\label{sec:prompt-for-generation}

\revise{The more detailed prompts during the conversation process are structured into 6 parts, including overall prompt, scenarios (S1, S2), user records, conversational partner's persona, reference of closeness, and chat history.}


\revise{Every conversational partner's persona stores the two dimensions influencing sentence suggestion generation: topic preferences and closeness level.}

\revise{In the overall prompt, we first instruct GPT to assume the role of an AAC tool and express in the text the social communication task and objectives. Next, we provide task details in S1 and S2 based on the two different communication scenarios: starting new conversations and responding to others. The user record refers to the relevant records that have been extracted in the previous step (Step 5 in Preprocessing). In the chat history, the user's conversation history with the current conversational partner is presented.}




\revise{In the prompt, the two important dimensions of social communication play a role in S1 and S2. Topic preferences serve as references when selecting records to start new conversation. During response generation, closeness becomes the key reference. Based on prior research~\cite{rocca1998relationship}, we define three levels of closeness—average, familiar, and very familiar, which influence conversational depth. These levels are included in the prompt to guide GPT’s sentence suggestions:}

\begin{itemize}
  \item \textbf{Average:} Use polite, courteous, and distant language, and do not discuss details in user records at all.
  \item \textbf{Familiar:} Use a small amount of detail in the desired content, but still maintain a certain sense of caution and distance, briefly discuss the details in personal records.
  \item \textbf{Very Familiar:} Discuss details in user records, and tend to ask the conversation partner more detailed questions about the selected user records.
\end{itemize}

\subsection{Methods of Customizing Generated Sentences}



\revise{To respect diverse user preferences and avoid potential bias from system suggestions, SocializeChat offers two flexible input methods. }

\revise{First, users can refine suggested sentences by selecting contextual adjustment tags generated based on conversation context in real time. For example, in the case of last case, the tags are: Agree, Disagree, Hesitant, and Considerate. If the user selects the "Disagree" tag, GPT modifies its responses based on the original reply and the "Disagree" tag to align with a disagreeable attitude. }

\revise{Second, full manual input empowers users to compose messages freely, without any system constraint. This manually input feature is fully compatible with iOS accessibility functions, supporting elderly users with impairments and maximizing user autonomy.}

\section{User Study}

\subsection{Procedure}


\paragraph{Step 1: Introduction to SocializeChat (10 min)}
Through storyboarding, we provided participants with an introduction to the usage of our product.

\paragraph{Step 2: Trying out the Tool (30 min)}

Participants first set the closeness level with their conversation partners and configured personal records. Our team assisted as needed, especially during personal record setup, encouraging participants to share rich, detailed entries. After setup, participants underwent a 10-minute familiarization phase.


In the formal test, each participant engaged in a 15-minute conversation via SocializeChat with a partner of their choice (a familiar person or an assigned one). Participants could start with system-suggested topics or respond to partner-initiated topics based on pre-recorded personal content. Although conversations began with these records, new topics often emerged naturally.

\paragraph{Step 3: Post-test survey (20 min)}
\revise{Given participants' varying speech impairments, the interview format was flexible. Most provided feedback through supporters, who paraphrased between us and the participants.}

\subsection{Recruitment and Data Collection}
We recruited 16 elderly Mandarin-speaking adults \revise{(7 males, 9 females; \(M = 66.77\), \(SD = 7.50\))} in mainland China with doctor-confirmed speech impairments (e.g., due to tooth loss or stroke; see Table \ref{tab:participant_full}). All had intact language comprehension but experienced difficulties with speech and writing. The study received ethics approval. 

\revise{Each chat round involved a user message and one reply. We measured the speaking rate in Chinese characters per minute (WPM) to assess efficiency. To evaluate sentence suggestion, we used a five-point scale from "completely mismatched" to "perfectly matched" and labeled topics~\cite{rocca1998relationship}. The SUS scale~\cite{bangor2008empirical} assessed usability, with questions read aloud and explained when needed.}




\section{Findings of User Study}
\subsection{Process of Conversations with SocializeChat}

\begin{table}[t]
\caption{Participant Profiles and Conversation Topic}
\label{tab:participant_full}
\centering
\scriptsize
\setlength{\tabcolsep}{1.5pt}  
\begin{tabular}{|p{0.05\linewidth}|p{0.05\linewidth}|p{0.1\linewidth}|p{0.1\linewidth}|p{0.2\linewidth}|p{0.4\linewidth}|}
\hline
\textbf{ID} & \textbf{Age} & \textbf{Non-verbal?} & \textbf{Non-motor?} & \textbf{Partner (Closeness)} & \textbf{Topics} \\
\hline
P1 & 75 & Full & Part. & Family (V.Fam.) & Weather, Daily activities, Hobbies, Future plans and goals \\
P2 & 55 & Smt. & Slight & Family (V.Fam.) & Daily activities, Hobbies \\
P3 & 72 & Smt. & No & Family (V.Fam.) & Shared past experiences, Politics, Social activities, Friends, Money \\
P4 & 73 & Full & Part. & Family (Fami.) & Daily activities, Other family members, Religion, Social activities \\
P5 & 62 & Smt. & No & Friend (Fami.) & Shared past experiences, Hobbies, Friends \\
P6 & 60 & No & No & Friend (Fami.) & Shared past experiences, Other family members, TV programs \\
P7 & 75 & Smt. & No & Friend (Avg.) & Daily activities, Hobbies, Shared past experiences \\
P8 & 69 & Smt. & No & Stranger (Avg.) & Weather, Daily activities, Hobbies, Jobs \\
P9 & 73 & No & No & Friend (Avg.) & Hobbies, Jobs \\
P10 & 76 & Smt. & No & Stranger (Avg.) & Hobbies, Jobs, Personal problem \\
P11 & 60 & Smt. & No & Stranger (Avg.) & Hobbies, Jobs, Daily activities \\
P12 & 77 & Full & Full & Stranger (Avg.) & Other family members, Weather \\
P13 & 57 & Smt. & Part. & Friend (Fami.) & Other family members, Shared past experiences, Sports \\
P14 & 68 & No & Part. & Stranger (Avg.) & Hobbies, Daily activities, Shared past experiences \\
P15 & 73 & Smt. & Part. & Stranger (Avg.) & Hobbies, Shared past experiences, Future plans and goals \\
P16 & 62 & Smt. & No & Stranger (Avg.) & Daily activities, Religion \\
\hline
\end{tabular}
\end{table}

During an average of 15.15 ($SD = 1.61$) minutes, participants had 16.06 ($SD=2.96$) average conversation rounds. Among the participants, P2 had the highest number of rounds with 22 conversations, while P7 and P10 had fewer rounds, each with only 12 conversations. \revise{Moreover, the average WPM is 51.99 ($SD=14.34$). SocializeChat allowed the participants to chat more efficiently with others. The average score on the SUS scale is 70.68, suggesting that SocializeChat is generally acceptable to the participants~\cite{bangor2008empirical}. }

\revise{Despite the relatively slow speaking rate, participants did not perceive the selection process as challenging. Most felt that the sentence suggestions aligned well with their personal thoughts and memory records. Among the 16 participants, 7 rated the suggestions as Perfectly Matched, 4 as Mostly Matched, 3 as Moderately Matched, and only 2 expressed any mismatch. These responses suggest that the system effectively leveraged user records to generate contextually appropriate suggestions, as also reflected by the frequency of customization ($M=2.0, SD=2.68$). The participants aslo expressed that the sentence suggestions presented captured the content they wanted to convey, and as a result, they made their choices without much hesitation. }

Conversation topics are diverse spanned 18 categories (Table~\ref{tab:participant_full}). Most common were daily activities and hobbies (8), followed by shared past experiences (7), family (4), friends and weather (3). Less frequent topics—each mentioned once or twice—included social life, future plans, money, religion, school, entertainment, jobs, politics, advice, sports, and personal issues.




\subsection{Leveraging Personal Memory Reords through Closeness-Aware Suggestions}

Most participants ($N=11$) emphasized the importance of reflecting interpersonal closeness in social conversations. They appreciated how SocializeChat adjusted sentence content based on relationship levels, often surfacing details from their prior records. This allowed users to express themselves in a manner aligned with how they normally communicate—sharing more openly with close family and being more reserved with strangers.

Sentence suggestions for closer relationships were perceived as richer and more emotionally appropriate. Participants found that these suggestions supported natural everyday interactions, such as sharing family routines or small joys. Suggestions grounded in personal experience enabled users to maintain familiarity and warmth in their conversations.

Topic-based suggestions were also appreciated for their ability to trigger relevant memory records. Participants felt that talking about shared interests helped reduce the cognitive load of initiating conversations and promoted meaningful engagement.

Even after the sessions ended, participants expressed a desire to continue. Several mentioned having more to say about the topics in the conversations or wished to share personal memories, like a recent grape picking (P11) or experience on rowing (P13). These examples show that memory-based conversation not only supported communication but also stimulated users’ motivation to connect.

\section{Discussion}


\subsection{Supporting Memory-Based Social Communication}

Participants appreciated how SocializeChat allowed them to express themselves in a way that reflected their real-life interactions. Sentence suggestions grounded in prior records and tailored by interpersonal closeness enabled users to share familiar stories, routines, and interests. This memory-based design not only made the system feel more relatable but also encouraged users to engage in longer and more meaningful conversations.

Even though the system’s speaking speed was slower than fluent speakers, most users did not perceive it as a barrier. Their attention was drawn instead to the alignment between sentence content and their social intentions. This suggests that, for users with rich personal histories, content relevance may outweigh speed in determining the usefulness of an AAC system.

\subsection{Balancing Control and Expressive Freedom}

User autonomy in communication remains a critical concern in AI-supported AAC systems~\cite{valencia2023less}. SocializeChat preserved this autonomy by allowing manual edits, contextual customization, and content generation based on users' own records. While LLMs occasionally generated content inconsistent with the user’s reality, participants adopted flexible strategies: some revised the output to better match their experiences, others accepted the suggestions as socially adequate. This highlights the importance of designing for both factual accuracy and emotional or relational appropriateness in memory-based conversations.

\subsection{Human-AI Co-Construction of Social Identity}

SocializeChat reflects a shift from full manual control to co-constructed expression, where users guide the AI using personal memories, preferences, and relationship context~\cite{guzman2020artificial}. This design echoes Peter Scott-Morgan's vision of AI-assisted communication as a form of social augmentation~\cite{aylett2022peter}, in contrast to Stephen Hawking's emphasis on precision and full control. Our findings suggest that older adults are willing to partially delegate expressive content when it helps them articulate meaningful, memory-anchored narratives. Future research should continue to investigate how AI systems can responsibly support such co-authored expressions in diverse communication scenarios.


\subsection{Limitation and Future Work}

While memory design often leverages multimedia resources~\cite{wang2024promoting, chan2021pablo, wei2023bridging}, future work should explore formats beyond text to better support AAC users’ multimodal communication needs. Our participants were elderly adults from Mainland China, and cultural and regional factors~\cite{hetzroni1996cultural} as well as \revise{LLM capabilities~\cite{xu2024exploringmultilingualconceptshuman}} may limit the generalizability of our findings. \revise{We did not include another group using traditional AAC tools due to fundamental differences in their design, but plan to conduct comparative studies in future work. Moreover, participants' varying digital literacy may also have influenced their efficiency and interaction with the system. Although not formally assessed, this factor should be considered in interpreting results and addressed in future recruitment and design. We also plan to incorporate GPT-based autocomplete to reduce input burden and better support elderly users.}

\section{Conclusion}

We presented SocializeChat, a GPT-based AAC tool designed to support social communication by drawing on older adults’ personal memories including records and partners' personas. By aligning sentence suggestions with familiar topics and interpersonal closeness, SocializeChat helped users express themselves in ways that felt authentic and socially appropriate. Our findings highlight the value of this memory-based design in AAC and suggest that older adults are open to co-constructing expressions with AI when the system respects their lived experience. We hope this work encourages future AAC research to further explore how memory, identity, and personalization can shape inclusive communication technologies.

\bibliographystyle{IEEEtran}  
\bibliography{refs}

\end{document}